\documentclass[%
 aip,
rsi,%
 amsmath,amssymb,
 reprint,%
]{revtex4-1}

\usepackage{xcolor}
\usepackage{graphicx}% Include figure files
\usepackage{dcolumn}% Align table columns on decimal point
\usepackage{bm}% bold math
\usepackage{nicefrac}
\newcommand{\n}[1]{\mathbf{#1}}

\begin{document}

\preprint{AIP/123-QED}

\title[Rodr\'{i}guez et al.]{Connection between quasi-symmetric magnetic fields and anisotropic pressure equilibria in fusion plasmas}% Force line breaks with \\

\author{E. Rodr\'{i}guez}
 \altaffiliation[Email: ]{eduardor@princeton.edu}%Lines break automatically or can be forced with \\
 \affiliation{ 
Department of Astrophysical Sciences, Princeton University, Princeton, NJ, 08543%\\This line break forced with \textbackslash\textbackslash
}
\affiliation{%
Princeton Plasma Physics Laboratory, Princeton, NJ, 08540%\\This line break forced% with \\
}%

\author{A. Bhattacharjee}
 \altaffiliation[Email: ]{amitava@princeton.edu}
 \affiliation{ 
Department of Astrophysical Sciences, Princeton University, Princeton, NJ, 08543%\\This line break forced with \textbackslash\textbackslash
}
\affiliation{%
Princeton Plasma Physics Laboratory, Princeton, NJ, 08540%\\This line break forced% with \\
}%

\date{\today}% It is always \today, today,

\begin{abstract}
The \textit{stellarator} as a concept of magnetic confinement fusion requires careful design to confine particles effectively. A design possibility is to equip the magnetic field with a property known as \textit{quasisymmetry}. Though it is generally believed that a steady-state quasisymmetric equilibrium can only be exact locally (unless the system has a direction of continuous symmetry such as the tokamak), we suggest in this work that a change in the equilibrium paradigm can ameliorate this limitation. We demonstrate that there exists a deep physical connection between quasisymmetry and magnetostatic equilibria with \textit{anisotropic} pressure, extending beyond the isotropic pressure equilibria commonly considered.
\end{abstract}

\maketitle

% Magnetic confinement is one of the many forms presently considered in the pursue of controlled thermonuclear fusion.\cite{huang2018,craxton2015,wurden2016} Stellarators\cite{spitzer1958}, three-dimensional generalisations of tokamaks, have recently seen a resurgence alongside the notion of quasisymmetry, which confers them superior confinement properties necessary to attain large plasma temperatures, while keeping useful 3D features. This field property is generally believed to be only achievable locally\cite{garrenboozer1991b} or in continuously symmetric cases. In this work we show through a constrained energy principle that equilibria with anisotropic pressure are a more natural form of equilibrium to quasisymmetric fields. This new perspective opens the door to a larger design space (including more global quasisymmetric configurations), as well as a deeper understanding on the implications and feasibility of quasisymmetry in practice.
Ever since Lyman-Spitzer invented the \textit{stellarator}\cite{spitzer1958}, this inherently three-dimensional, steady-state concept for confining fusion plasmas magnetically has held the promise to be an attractive alternative to tokamaks, which are prone to disruptive instabilities. Unlike tokamaks, for which axisymmetry provides good confinement of particles and energy, stellarators \textit{rely} on symmetry breaking to realize the magnetic field. Over the last few decades, the discovery of \textit{hidden symmetries} has led to a renaissance of the stellarator concept. A prominent example of a hidden symmetry is \textit{quasisymmetry} (QS)\cite{boozer1983,nuhren1988,Helander2014,rodriguez2020a,burby2020}, which has guided numerous designs and experiments\cite{hsx1995,zarnstorff2001,drevlak2013,henneberg2019,bader2019}. \par
We define \textit{quasisymmetry} as the minimal property of a magnetic field that provides the dynamics of charged particles with an approximately conserved momentum.\cite{rodriguez2020a,burby2020} This conservation prevents (as \textit{Tamm's theorem} does in an axisymmetric device) particles from drifting away from the stellarator. By Noether's theorem, this conservation should be conjugate to a symmetry of the magnetic field. A quasisymmetric configuration bears that symmetry on the magnitude of the magnetic field, $|\n{B}|$, but does not in $\mathbf{B}$. \par
The implications of such symmetry had long been recognised\cite{boozer1983,nuhren1988,Helander2014} in the context of magnetohydrostatic equilibrium with isotropic pressure, $p$ (referred to hereafter as MS equilibrium). Only recently\cite{rodriguez2020a,burby2020} we have been able to formulate the concept of QS based entirely on single-particle orbits. separating it from assumptions regarding equilibria. Doing so allows for a general and succinct definition of QS as a magnetic field with well-defined flux surfaces (labelled by the variable $\psi$) for which $f_T=\nabla\psi\cdot\nabla B\times\nabla(\n{B}\cdot\nabla B)=0$.\cite{rodriguez2020a,Helander2014} We call this the \textit{triple vector formulation} of QS.
\par
Liberated from the particular form of MS equilibria, we ask what type of equilibrium is natural for QS. The traditional approach is to think of MS equilbria as states of minumum energy to which toroidal plasmas relax when their evolution is governed by ideal magnetohydrogynamic (MHD) laws (with some measure of flow damping). This classical formulation is due to Kruskal and Kulsrud\cite{kruskuls1958}, who elegantly presented the problem through a variational (energy) principle. Define the energy functional,
\begin{equation}
    \mathcal{W}_0=\int_\mathcal{V}\left(\frac{B^2}{2}+\frac{p}{\gamma-1}\right)\mathrm{d}\tau,
\end{equation}
where $\mathcal{V}$ is a fixed toroidal volume with boundary $\partial\mathcal{V}$ as a flux surface, and $\gamma$ is the adiabatic coefficient. The extrema of $\mathcal{W}_0$ are precisely MS equilibria $\n{j}\times\n{B}=\nabla p$, where $\n{j}$ is the plasma current density.\cite{bhattacharjee1980}  \par
% where $\mathrm{d}\tau$ is the infinitesimal volume element. The equilibrium then follows from the Euler-Lagrange (EL) equations associated to $\mathcal{W}_0$. The variation $\delta\mathcal{W}_0$ may be evaluated by including ideal MHD equations through virtual displacements $\xi$.\cite{bhattacharjee1982} Variations of the fields are then $\delta\n{B}=\nabla\times(\xi\times\n{B})$ and $\delta p=-\xi\cdot\nabla p-\gamma p\nabla\cdot\xi$. Requiring the extrema $\delta\mathcal{W}_0=0$ to hold for an arbitrary $\xi$, the EL equation is $\n{j}\times\n{B}=\nabla p$; i.e., magnetohydrostatic balance with isotropic pressure (MS). \par
This energy perspective on equilibrium presents MS as a natural state for a toroidal plasma. However, this does not guarantee the resulting equilibrium to be quasisymmetric, and it will generally not be so. Our challenge is to enforce the constraint of QS in the formulation of Kruskal and Kulsrud to understand what the equilibria for a QS field would be. \par
\begin{figure}
    \centering
    \includegraphics[]{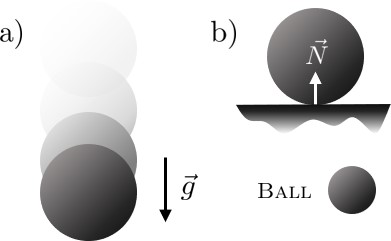}
    \caption{\textbf{Classical falling ball analogue.} Schematic for a ball under gravity which tries to minimise energy. a) Falling ball under the effect of gravity, $\vec{g}$, which will do so indefinitely. b) For the ball to stand at a certain height, the presence of a normal force, $\vec{N}$, is needed.}
    \label{fig:ballFall}
\end{figure}

We draw here from intuition developed through a mechanical analogy\cite{gelfand2000,arnold2006,marsden2001}. As a simple reference example take a ball under the influence of gravity which is forced to rest on the ground (see Fig.~\ref{fig:ballFall}). To formulate constrained problems of this and a more complex nature, we define i) an action functional $\mathcal{S}[q_i,\dot{q}_i]=\int\mathcal{L}(t,q_i,\dot{q}_i)\mathrm{d}t$, where $\mathcal{L}$ is the Lagrangian and $q_i$ are generalised coordinates, and ii) the corresponding \textit{holonomic} constraints\cite{goldstein2002,arnold2006} $f_j(q_i)=0$ for $j=1,\dots,m$. These two pieces can be accomodated through the addition of a Lagrange multiplier $\lambda_j(t)$, to give a modified constrained functional $\mathcal{S}_\lambda=\int\left(\mathcal{L}+\sum_j \lambda_j(t)f_j\right)\mathrm{d}t$. The resultant modified Euler-Lagrange equations, $\mathrm{d}/\mathrm{d}t(\partial\mathcal{L}/\partial \dot{q}_i)-\partial\mathcal{L}/\partial q_i=Q_i$, include generalised forces\cite{goldstein2002} $Q_i=\sum_{j=1}^m\lambda_j(t)\partial f_j/\partial q_i$. These additional forces are needed to guarantee that the dynamics of the system will not violate the imposed constraint. In the falling ball problem, a normal force is necessary to prevent the ball from continuing its fall. So if the ball is wanted at a particular elevation, an external force is required. \par
We now extend this elementary picture to the problem of imposing QS into the energy functional governing plasma relaxation. The relevant equilibrium-independent constraint to impose QS is $f_T=0$, a holonomic constraint in the context of continuum mechanics\cite{marsden2001}. Using a space dependant Lagrange multiplier $\lambda(\n{r})$,\cite{marsden2001,gelfand2000,arnold2006} we define the constrained form of the energy principle to be,
\begin{equation}
    \mathcal{W}_\lambda=\int\left(\frac{B^2}{2}+\frac{p}{\gamma-1}+\lambda(\n{r})f_T\right)\mathrm{d}\tau. \label{eqn:Wlambda}
\end{equation}
As a result, we expect to find a new generalised force, needed to prevent the system from \textit{falling} to an unconstrained minimum-energy state without QS. The extrema of $\mathcal{W}_\lambda$ can be shown to give the following Euler-Lagrange equation,
\begin{equation}
    \n{j}\times\n{B}-\nabla p = T_1\nabla\psi+\n{B}\times\nabla\times\left[T_2\n{b}-T_3\nabla B+(\n{B}\cdot\nabla T_3)\n{b}\right], \label{eqn:ELWlambda}
\end{equation} 
where $T_1=\nabla\lambda\cdot\nabla B\times\nabla(\n{B}\cdot\nabla B)$, $T_2=\nabla\lambda\cdot\nabla\psi\times\nabla(\n{B}\cdot\nabla B)$ and $T_3=\nabla\lambda\cdot\nabla\psi\times\nabla B$. The left hand side of Eq.~(\ref{eqn:ELWlambda}) has the form of MS equilibrium, which leaves the right-hand-side as the \textit{generalised force}. The presence of this force is necessary to maintain QS and prevent the system from relaxing to the minimum-energy MS equilibrium. The Lagrange multiplier, determined by the triple vector constraint $f_T=0$, is a local measure of the cost of enforcing QS.  \par
The remarkable feature of Eq.~(\ref{eqn:ELWlambda}), despite the seemingly artificial form of the forcing term, is that it can be recast into the form,
\begin{equation}
    (1-\Delta)\n{j}\times\n{B}=\nabla p_\perp+(\n{B}\cdot\nabla\Delta)\n{B}+\Delta\nabla\left(\frac{B^2}{2}\right), \label{eqn:anisFB}
\end{equation}
where $ p_\parallel=p-T_3(\n{B}\cdot\nabla B)$, $p_\perp=(p+BT_2)+\n{B}\cdot\nabla\left(T_3/B\right)B^2$ and $\Delta=(p_\parallel-p_\perp)/B^2$. Equation~(\ref{eqn:anisFB}) is precisely the equation for the equilibrium of a plasma with a diagonal anisotropic pressure tensor $\Pi=(p_\parallel-p_\perp)\n{b}\n{b}+p_\perp\mathbb{I}$, where $\mathbb{I}$ is the unit dyad and $\n{b}=\n{B}/|\n{B}|$. The system does bring in, unexpectedly but naturally,  anisotropic pressure into the relaxed equilibrium state, establishing a deep connection between QS and MHD equilibria with anisotropic pressure. \par
The form of the anisotropy found through the variational process is not arbitrary. In fact, taking $\lambda$ to be a single-valued function with no special symmetry property, the forms of the pressure from the Euler-Lagrange equation must obey the relations
\begin{equation}
    \oint_{\psi,B}\Delta\mathrm{d}\alpha=0, \label{eqn:relAnis}
\end{equation}
and $p(\psi)=\oint_{\psi,B}p_\parallel\mathrm{d}\alpha/2\pi(N-\iota)$. Here $\alpha$ is a field line label (with integrals being taken along the symmetry direction of $|\n{B}|$), $\iota$ is the rotational transform of the field, and $N$ represents the pitch of the constant-$B$ streamlines in generalised Boozer coordinates\cite{garrenboozer1991b,rodriguez2020i}. Equation~(\ref{eqn:relAnis}) represents a pressure anisotropy close to the isotropic $\Delta=0$ form, but which generally departs through a field line dependence because of QS. On the other hand, the average of the perpendicular and parallel pressure yield $p(\psi)$, the scalar pressure as introduced in Eq.~(\ref{eqn:Wlambda}). These forms are consistent with MS, in the sense that the latter is a subset of the former. \par
%  It might seem somehow contradictory to have both $p$ and anisotropic pieces at the same time in the problem. One may nevertheless interpret $p$ simply as the plasma internal energy content, {\color{red} or note that the quasisymmetric forcing piece was a magnetic piece and thus an identical term would arise in a CGL or ANIMEC-like energy principle}. \par
\begin{figure*}
    \centering
    \includegraphics[]{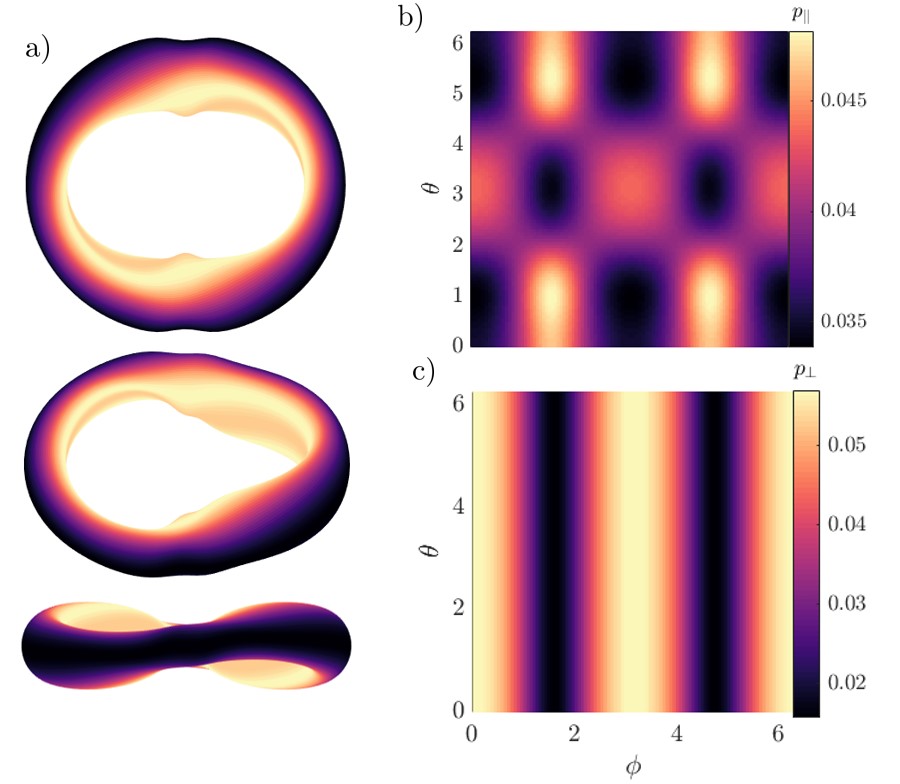}
    \caption{\textbf{Quasisymmetric stellarator solution with anisotropic pressure.} Quasisymmetric stellarator equilibrium with anisotropic pressure of the form obtained from the variational principle Eq.~(\ref{eqn:relAnis}) by near-axis expansion through second order. The configuration is a modified version of the stellarator in Sec.~5.3 in [\onlinecite{landreman2019}] to comply with the constrained variational equilibrium here. a) Projections of the stellarator 3D shape at three angles, with colormap denoting magnetic field magnitude strength. b) Parallel pressure on the toroidal angle, $\phi$, and poloidal angle, $\theta$ space on the shown surface in a). c) Perpendicular pressure on the $(\phi,\theta)$ plane. $(\phi,\theta)$ are magnetic flux angular coordinates. Pressures are given in reference to magnetic pressure.}
    \label{fig:anisStellQS}
\end{figure*}
The appearence of this form of anisotropic pressure in the equilibrium of the problem opens the door to two lines of interpretation. The first one is to understand this form of equilibrium, namely Eq.~(\ref{eqn:anisFB}), as a truly \textit{physical} equilibrium, which can be realized in practice. The treatment given in this paper suggests that MHD equilibria with anisotropic pressure are more suited to configurations that are quasisymmetric everywhere, and are thus of fundamental as well as practical interest. For this equilibrium to be realistic, the macroscopic results  obtained here need to be reconciled with kinetic theory. Pressure has a very specific meaning kinetically as the centred second moment of the distribution function describing the plasma in phase space (which, for instance, requires $p_\parallel,~p_\perp\geq0$). Different forms of the distribution function at different time scales and orderings will have different implications on the allowable forms and sizes of $(p_\parallel,p_\perp)$. A kinetic study that analyses in what scenarios is the constrained variational equilibrium a physically achievable solution is left for a future publication. 
% Nevertheless, it is important to bear in mind that, formally speaking, one expects to always find anisotropic pieces from the kinetic treatment, even though these might be small in the appropriate ordering.
\par
The second perspective on the anisotropic equilibrium obtained here is to view it as a formal tool by which we are able to extend the space of quasisymmetric solutions. The form of Eq.~(\ref{eqn:anisFB}) is formally \textit{very} different from the MS equilibrium equations, and through its link to QS, opens up a more convenient space in which to examine the question of globally quasisymmetric solutions. This space described by Eq.~(\ref{eqn:anisFB}) remains formally different from MS even as $\Delta\rightarrow0$, possibly including solutions close to isotropy, but which lie outside the MS space of solutions. We do not attempt this here, but remark that our proposition is qualitatively consistent with the Constantin-Drivas-Ginsberg (CDG) theorem\cite{constantin2021}, which proves existence of quasisymmetric solutions under some restrictive conditions in the presence of a residual forcing.  \par
Thus, we conclude that quasisymmetric equilibrium solutions with anisotropic pressure are of fundamantal and practical interest. One way to obtain such solutions, numerically, would be to use Eq.~(\ref{eqn:Wlambda}) to formulate a numerical variational or optimisation problem. Such approaches to equilibrium solutions have proved to be of great practical use\cite{hirshman1983,bhattacharjee1984,cooper1992}, with representative codes such as VMEC\cite{hirshman1983} and ANIMEC\cite{cooper1992}, which could be modified to incorporate the QS constraint. However, we consider an alternative approach here, which involves the so-called \textit{near-axis expansion}\cite{rodriguez2020i,landreman2018a,garrenboozer1991b}. \par
At the heart of this method is to expand solutions and governing equations in powers of the distance to the magnetic axis (see Appendix and referenced work for more details), and solve the resulting equations order by order. When MS equilibria are considered, this approach breaks down due to what is now known as the \textit{Garren-Boozer overdetermination problem}\cite{garrenboozer1991b}. In brief, Garren and Boozer showed that the process of expansion for quasisymmetric solutions leads to an overdetermined system of equations. This conundrum has been widely interpreted to mean that global quasisymmetric solutions do not exist but in cases of continuous symmetry such as axisymmetry or helical symmetry. However, following [\onlinecite{rodriguez2020i}], we have demonstrated that the Garren-Boozer overdetermination problem can be resolved when solutions to Eq.~(\ref{eqn:anisFB}) are considered. In Fig.~\ref{fig:anisStellQS} we present, for the first time (previously only done for circular axes\cite{rodriguez2020ii}), a quasisymmetric equilibrium solution exact through second order in the expansion. This stellarator configuration was suggested in [\onlinecite{landreman2019}], where the MS limitations prevented quasisymmetry from being achieved to second order. These numerical solutions are further evidence of the deep connection between anisotropic pressure and QS. \par
In summary, we demonstrate that there exists a deep connection between quasisymmetric fields in equilibria and anisotropic pressure. We do so by presenting a variational principle in which the energy is extremised subject to the QS constraint, yielding a special realisation of equilibria with anisotropic pressure. These results prompt a change in the equilibrium paradigm, pointing to the possibility of globally quasisymmetric solutions. In this paper, we illustrate this by constructing explicit numerical higher-order quasisymmetric configurations through near-axis expansions.
\par
We conclude by thanking P. Helander, E. Paul and W. Sengupta for stimulating discussions. This research was supported by a grant from the Simons Foundation/SFARI (560651, AB) and DoE Grant No. DE-AC02-09CH11466.

\appendix
% \section{Extremising the functionals}
% In order to extremise the unconstrained or constrained anergy functionals presented in this Letter, it is important to keep in mind that not all field variations are allowed. In fact, only variations that abide by the laws of ideal MHD are permitted. A straightforward way to formally incorporate this constraint is to introduce the concept of virtual displacements $\xi(\mathbf{r},t)$.\cite{bhattacharjee1980} When we vary the fields, we shall do so by virtually displacing plasma parcels by some amount $\xi$. Then, from the MHD equations it follows that fields must vary according to $\delta\n{B}=\nabla\times(\xi\times\n{B})$ (induction equation) and $\delta p=-\xi\cdot\nabla p-\gamma p \nabla\cdot\xi$ (adiabatic relation). Taking the variation $\delta\mathcal{W}$, and integrating by parts, one can recast the first variation of the functionals in the general form,
% \begin{equation}
%     \delta\mathcal{W}=\int\xi\cdot\left(\dots\right)\mathrm{d}\tau. \label{eqn:genFormVar}
% \end{equation}
% For the functional to be an extrema, this variation must vanish \textit{for all} $\xi$. It then follows that the term in brackets in the integrand of Eq.~(\ref{eqn:genFormVar}) abstractly referenced by $(\dots)$ must vanish. This vector equation describes equilibria: it gives MS equilibria for $\mathcal{W}_0$ and Eq.~(\ref{eqn:ELWlambda}) for $\mathcal{W}_\lambda$. 

\section{Constructing solutions by near-axis expansion} \label{sec:appNAE}
The main idea behind the construction of solutions by near-axis expansion is straightforward. Instead of attempting to find global solutions to the set of governing equations (here a form of equilibrium and the QS condition), we instead expand these perturbatively in powers of the distance from the magnetic axis. This will generally lead to a hierarchy of simpler equations that need to be solved order by order. The details of such procedure had been provided for the case of MS equilibria by [\onlinecite{garrenboozer1991a,garrenboozer1991b,landreman2018a,landreman2019}] and only recently for more general forms of equilibria by [\onlinecite{rodriguez2020i,rodriguez2020ii}]. In the near-axis formulation, different configurations are described by a different set of constant parameters and magnetic axis shapes, some of which are free and some of which need to be obtained self-consistently\cite{landreman2018a,rodriguez2020i}. \par
To construct solutions such as that shown in Fig.~\ref{fig:anisStellQS} of this paper, we follow the general scheme introduced in [\onlinecite{rodriguez2020i}] applied to an equilibrium of the form of Eq.~(\ref{eqn:anisFB}) and Eq.~(\ref{eqn:relAnis}). No prior such numerical solution exists, as numerical solutions had previously only been provided for the simplest of shapes in [\onlinecite{rodriguez2020ii}] or MS equilibria [\onlinecite{landreman2019}]. A set of equations analogous, albeit more complex, to those in [\onlinecite{rodriguez2020ii}] need to be solved here. To obtain and solve such equations, we follow the methodology and steps described in [\onlinecite{rodriguez2020i}] and [\onlinecite{rodriguez2020ii}]. We shall not reproduce those equations here.  \par
Given that some of the parameters describing the solution are free, for Fig.~\ref{fig:anisStellQS} we have opted for the example provided in Sec.~5.3 of [\onlinecite{landreman2019}] as a starting point. Note that to find an appropriate quasisymmetric solution to second order some of the parameters need to be modified in a self-consistent way (example of which is parameter $\bar{\Delta}_{20}$ in [\onlinecite{rodriguez2020ii}]). In addition, in the present scenario Eq.~(\ref{eqn:relAnis}) imposes an additional constraint on parameters; in particular, it requires $\oint_{\psi,B}\Delta_{nm}\mathrm{d}\phi=0$ and similarly for $p_{nm}$ with $m\neq0$, where the closed integrals are at constant $\psi$ and $|\mathbf{B}|$. The search for a consistent set of parameters is run as an optimisation problem. As a result of the approach, the parameters describing Fig.~\ref{fig:anisStellQS} are: $\sigma(0)=1.01\times10^{-4}$, $\bar{B}_{\theta 20}=2.8546$, $\eta/\sqrt{2}=0.95$, $p_0=0.08$, $\Delta_0=0$, $B_{22}^C=5.51$, $B_{22}^S=0$, $B_{20}=-3.69$, $B_{31}^C=0.01$, $B_{31}^S=0.01$, $R_\mathrm{ax}=1+0.09\cos2\phi$, $Z_\mathrm{ax}=-0.09\sin2\phi$, $B_{\alpha0}=1.02$, $B_{\alpha1}=2.04$, $\epsilon=0.1414$. To compare these to [\onlinecite{landreman2019}], one must be careful, as here parameters have been defined as in [\onlinecite{rodriguez2020i}]. To go back and forth between this form and that of [\onlinecite{landreman2019}] (which we denote by superscript L), and taking $B_0=1$ for simplicity, the main transformations are 
\begin{gather*}
    \eta^\mathrm{L}=\frac{\eta}{\sqrt{2}}, \\
    B_{20}^\mathrm{L}=\frac{3}{8}\eta^2-\frac{B_{20}}{4}, \\
    B_{22c}^\mathrm{L}=\frac{3}{8}\eta^2-\frac{B_{22}^C}{4}, \\
    B_{22s}^\mathrm{L}=-\frac{B_{20}}{4}, \\
    I_2^\mathrm{L}=\frac{\bar{B}_{\theta20}}{2}, \\
    p_2^\mathrm{L}=\frac{p_{20}}{2}, \\
    \sigma(0)^\mathrm{L}=\sigma(0).       
\end{gather*}

\hrulefill
\bibliography{variationalPRL}% Produces the bibliography via BibTeX.

\end{document}